# Fundamental properties of beam-splitters in classical and quantum optics

Masud Mansuripur and Ewan M. Wright

James C. Wyant College of Optical Sciences, The University of Arizona, Tucson, AZ 85721

**Abstract**. A lossless beam-splitter has certain (complex-valued) probability amplitudes for sending an incoming photon into one of two possible directions. We use elementary laws of classical and quantum optics to obtain general relations among the magnitudes and phases of these probability amplitudes. Proceeding to examine a pair of (nearly) single-mode wavepackets in the number-states $|n_1\rangle$ and $|n_2\rangle$ that simultaneously arrive at the splitter's input ports, we find the distribution of photon-number states at the output ports using an argument inspired by Feynman's scattering analysis of indistinguishable Bose particles. The result thus obtained coincides with that of the standard quantum-optical treatment of beam-splitters via annihilation and creation operators $\hat{a}$ and $\hat{a}^\dagger$. A simple application of the Feynman method provides a form of justification for the Bose enhancement implicit in the well-known formulas $\hat{a}|n\rangle = \sqrt{n}|n-1\rangle$ and $\hat{a}^\dagger|n\rangle = \sqrt{n+1}|n+1\rangle$.

**1. Introduction**. In quantum optics, as in classical optics, beam-splitters play an important role in many experimental settings.[1,2] Typically, a lossless beam-splitter has two input ports (1 and 2) as well as two output ports (3 and 4). A well-collimated wavepacket propagating in free space and arriving at one of the input ports can, to good approximation, be said to have frequency $\omega$, wave-vector $\boldsymbol{k} = (\omega/c)\hat{\boldsymbol{\kappa}}$, and polarization state $\hat{\boldsymbol{e}}$.[3,4] Upon entering an input port, the incident packet splits into a pair of wavepackets $(\omega, \boldsymbol{k}_3, \hat{\boldsymbol{e}})$ and $(\omega, \boldsymbol{k}_4, \hat{\boldsymbol{e}})$ that emerge from ports 3 and 4 along the directions $\hat{\boldsymbol{\kappa}}_3$ and $\hat{\boldsymbol{\kappa}}_4$, respectively. In classical optics, one states that the $E$-field amplitude of the incoming packet, either $E_1\hat{\boldsymbol{e}}$ or $E_2\hat{\boldsymbol{e}}$, divides between those of the outgoing packets $E_3\hat{\boldsymbol{e}}$ and $E_4\hat{\boldsymbol{e}}$. The various output-to-input ratios $E_{\text{out}}/E_{\text{in}}$ (generally, four complex numbers) then serve to specify the splitter's operational characteristics. In practice, beam-splitters are often constructed in the form of multilayer dielectric stacks, in which case their characteristic output-to-input amplitude ratios are referred to as their Fresnel reflection and transmission coefficients.[5-9] In this paper, we shall refer to the various $E_{\text{out}}/E_{\text{in}}$ ratios as the Fresnel coefficients, even though the beam-splitter could have a different physical structure, such as a (lossless) diffraction grating.

The classical output-to-input amplitude ratios $E_{\text{out}}/E_{\text{in}}$ also serve as the (complex) probability amplitudes for single-mode wavepackets $(\omega, \boldsymbol{k}_{\text{in}}, \hat{\boldsymbol{e}})$ occupied by the single-photon-number-state $|1\rangle$ that enter an input port and then emerge from an output port as a single-mode packet $(\omega, \boldsymbol{k}_{\text{out}}, \hat{\boldsymbol{e}})$ in the number state $|1\rangle$.[1,2] In Sec.2, invoking the principle of energy conservation (or, equivalently, photon-number conservation) in addition to the classical time-reversal symmetry of lossless splitters,[6,7] we derive general relations among the magnitudes and phases of the various $E_{\text{out}}/E_{\text{in}}$ ratios (i.e., Fresnel coefficients) corresponding to general-purpose lossless beam-splitters.

The behavior of a beam-splitter becomes much more complex (and far more interesting) in the quantum regime where a multi-photon packet $(\omega, \boldsymbol{k}_1, \hat{\boldsymbol{e}})$ in the number-state $|n_1\rangle$ arrives at port 1 of the splitter, while a second packet $(\omega, \boldsymbol{k}_2, \hat{\boldsymbol{e}})$ in the number-state $|n_2\rangle$ simultaneously arrives at its port 2.[10] In Sec.3, we derive the distribution of photon-number states in output ports 3 and 4 using an argument akin to the one used by Richard Feynman in his scattering analysis of indistinguishable Bose particles.[11] In a nutshell, Feynman asserts that the photon (a Bose particle) should be assumed to take all the allowed paths through a system, each path having its own probability amplitude—a complex number. When the various paths taken by the photon are physically indistinguishable, one must add all the probability amplitudes that lead from a specific initial condition to a specific final condition along different paths, the goal being to find the overall probability amplitude of the corresponding event. The probability of occurrence of the event is then the squared absolute value of the probability amplitude thus computed. The results of our analysis in Sec.3 will be seen to be in complete accord with the standard quantum-optical treatment of beam-splitters using the annihilation and creation operators $\hat{a}$ and $\hat{a}^\dagger$, as explained in Sec.4.

With its input in the $|n\rangle_1|0\rangle_2$ state, the splitter resembles a single-photon annihilator in the weak-reflectivity regime when the output state approaches $|1\rangle_3|n-1\rangle_4$. Similarly, when the input state is $|n\rangle_1|1\rangle_2$, the splitter acts as a single-photon creator in the limit when the state of the output approaches $|0\rangle_3|n+1\rangle_4$. These cases are briefly discussed in the examples presented at the end of Sec.4, where we argue that the well-known but often mysterious formulas $\hat{a}|n\rangle = \sqrt{n}|n-1\rangle$ and $\hat{a}^\dagger|n\rangle = \sqrt{n+1}|n+1\rangle$ can be justified even more directly by a straightforward application of the Feynman method. The paper closes with a brief summary and a few concluding remarks in Sec.5.

**2. Characteristics of beam-splitters**. Consider a transparent (i.e., non-absorbing) beam-splitter placed in a Michelson interferometer,[5] as shown in Fig.1(a). The Fresnel reflection and transmission coefficients at the point of initial incidence are $\rho = |\rho|e^{i\varphi_\rho}$ and $\tau = |\tau|e^{i\varphi_\tau}$. Upon returning from the first mirror, the light packet encounters a splitter whose Fresnel coefficients are $\rho' = |\rho'|e^{i\varphi'_\rho}$ and $\tau' = |\tau'|e^{i\varphi'_\tau}$; similarly, the packet returning from the second mirror (i.e., at the backside of the splitter) sees the coefficients $\rho'' = |\rho''|e^{i\varphi''_\rho}$ and $\tau'' = |\tau''|e^{i\varphi''_\tau}$; see Fig.1(b). Conservation of photon number (or energy) now demands that $|\rho|^2 + |\tau|^2 = 1$, $|\rho'|^2 + |\tau'|^2 = 1$, and $|\rho''|^2 + |\tau''|^2 = 1$. (Note that the interferometer is *not* required to have perpendicular arms when a general splitter is used.)

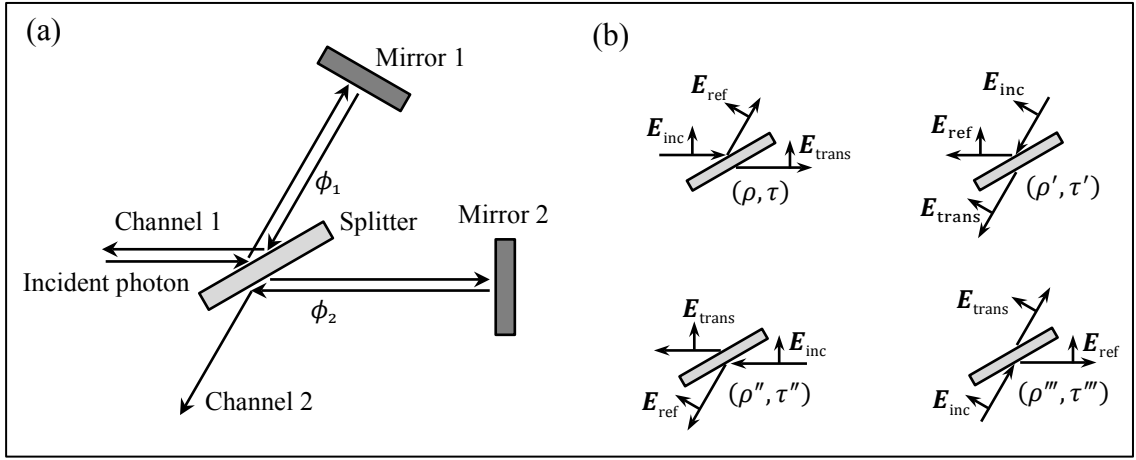

**Fig. 1**. (a) The beam-splitter used in a Michelson interferometer provides two possible paths for an incoming photon, one that allows the photon to be reflected back from Mirror 1 (roundtrip phase = $\phi_1$), and another that returns the photon after reflection from Mirror 2 (roundtrip phase = $\phi_2$). The returning photon has a probability amplitude $\psi_1$ for emerging from Channel 1 and a different amplitude $\psi_2$ for emerging from Channel 2. (b) Given a specific angle of incidence and a specific state of linear polarization, the splitter's reflection and transmission coefficients $\rho = E_{\text{ref}}/E_{\text{inc}}$ and $\tau = E_{\text{trans}}/E_{\text{inc}}$ represent the complex amplitudes of the reflected and transmitted $E$-fields corresponding to a unit-amplitude incident plane-wave. In general, the magnitudes and phases of $\rho$ and $\tau$ could depend on the side of the splitter from which the incident beam arrives, hence the designations $(\rho,\tau)$, $(\rho',\tau')$, $(\rho'',\tau'')$, $(\rho''',\tau''')$ for the four possible incidence geometries.

Let $\phi_1$ and $\phi_2$ denote the round-trip phase shifts accumulated along the two arms of the interferometer depicted in Fig.1(a). The probability amplitudes $\psi_1$ and $\psi_2$ for a single incident photon to emerge from channel 1 or channel 2 are readily found to be

$$\psi_1 = \rho\rho' e^{i\phi_1} + \tau\tau'' e^{i\phi_2}, \tag{1}$$

$$\psi_2 = \rho\tau' e^{i\phi_1} + \tau\rho'' e^{i\phi_2}. \tag{2}$$

Here we are using the probability amplitudes $\psi_1$ and $\psi_2$ for the emergence of a single photon from channel 1 or 2 instead of the returning $E$-field amplitudes $E_1 \hat{e}$ and $E_2 \hat{e}$ typically used in classical optics for a unit-amplitude incident $E$-field $\hat{e}$. Given that the overall probability of the photon emerging from one or the other channel should equal 1 (i.e., $|\psi_1|^2 + |\psi_2|^2 = 1$), we will have

$$|\psi_1|^2 + |\psi_2|^2 - 1 = |\rho\rho'|^2 + |\tau\tau''|^2 + 2|\rho\rho'\tau\tau''|\cos(\varphi_\rho + \varphi'_\rho - \varphi_\tau - \varphi''_\tau + \phi_1 - \phi_2)$$

$$+ |\rho\tau'|^2 + |\tau\rho''|^2 + 2|\rho\tau'\tau\rho''|\cos(\varphi_\rho + \varphi'_\tau - \varphi_\tau - \varphi''_\rho + \phi_1 - \phi_2) - 1$$

$$= |\rho|^2(|\rho'|^2 + |\tau'|^2) + |\tau|^2(|\rho''|^2 + |\tau''|^2) - 1$$

$$+ 2|\rho\tau|[|\rho'\tau''|\cos(\varphi_\rho + \varphi'_\rho - \varphi_\tau - \varphi''_\tau + \Delta\phi) + |\rho''\tau'|\cos(\varphi_\rho + \varphi'_\tau - \varphi_\tau - \varphi''_\rho + \Delta\phi)]$$

$$= 2|\rho\tau|[|\rho'\tau''|\cos(\varphi_\rho + \varphi'_\rho - \varphi_\tau - \varphi''_\tau + \Delta\phi) + |\rho''\tau'|\cos(\varphi_\rho + \varphi'_\tau - \varphi_\tau - \varphi''_\rho + \Delta\phi)] = 0. \quad (3)$$

The necessary and sufficient conditions for the satisfaction of Eq.(3) are given below; any other choice of the phase angles would make the required relation between $|\rho'\tau''|$ and $|\rho''\tau'|$ dependent on $\Delta\phi = \phi_1 - \phi_2$, an arbitrary parameter, which is inadmissible because the intrinsic properties of a beam-splitter should *not* depend on its distance from the interferometer's mirrors.

$$\varphi_\rho + \varphi'_\rho - \varphi_\tau - \varphi''_\tau = \varphi_\rho + \varphi'_\tau - \varphi_\tau - \varphi''_\rho \pm \pi$$

$$\rightarrow \boxed{(\varphi'_\rho + \varphi''_\rho) = (\varphi'_\tau + \varphi''_\tau) \pm \pi.} \quad (4)$$

The above choice of the phase angles then necessitates that $|\rho'\tau''|$ and $|\rho''\tau'|$ be equal; that is,

$$|\rho'\tau''| = |\rho''\tau'| \rightarrow |\rho'|^2(1 - |\rho''|^2) = |\rho''|^2(1 - |\rho'|^2)$$

$$\rightarrow \boxed{|\rho'| = |\rho''| \rightarrow |\tau'| = |\tau''|.} \quad (5)$$

The bottom line is that the splitter must have identical magnitudes $|\rho'|$ and $|\rho''|$ for the Fresnel reflection coefficients from its opposite sides and, similarly, identical magnitudes $|\tau'|$ and $|\tau''|$ of its corresponding transmission coefficients. The phase angles of $\rho'$ and $\rho''$, however, need not be the same, nor do the phase angles of $\tau'$ and $\tau''$, although $\varphi'_\rho + \varphi''_\rho$ must differ from $\varphi'_\tau + \varphi''_\tau$ by $\pi$.

Additional restrictions on the various Fresnel coefficients are imposed by the requirements of time-reversal symmetry.[6,7] Assuming the two mirrors of the Michelson interferometer are replaced by phase-conjugate mirrors,[12] we must have

$$\rho^*\rho' + \tau^*\tau'' = 1, \quad (6)$$

$$\rho^*\tau' + \tau^*\rho'' = 0. \quad (7)$$

A phase-conjugate mirror reflects an incident beam with its $E$-field magnitude intact but with the sign of its $E$-field phase reversed, causing the beam to retrace its path in space as though the arrow of time had been reversed (i.e., the time coordinate $t \rightarrow -t$). In the case of the Michelson interferometer depicted in Fig.1(a), replacing Mirrors 1 and 2 with phase-conjugate mirrors causes the reflected beams to return to the beam-splitter with the complex conjugates of their initial (complex) amplitudes, namely, $\rho^*$ returning from Mirror 1 and $\tau^*$ returning from Mirror 2. Since the beams are now retracing their original paths, they must combine at the splitter to reconstruct the incident beam in Channel 1 — albeit phase-conjugated and propagating backwards — in accordance with Eq.(6). There should be no (phase-conjugated) beam returning through Channel 2, since nothing had initially entered through this channel. The absence of a back-propagating beam in Channel 2 is now encoded in Eq.(7).

Multiplying Eq.(6) by $\tau'$ and Eq.(7) by $\rho'$, then subtracting the second equation from the first, we find

$$\tau^*\tau''\tau' - \tau^*\rho''\rho' = \tau'$$

$$\rightarrow \quad \tau^*\{|\rho'|^2 \exp[i(\varphi'_\rho + \varphi''_\rho)] - |\tau'|^2 \exp[i(\varphi'_\tau + \varphi''_\tau)]\} = -\tau'$$

$$\rightarrow \quad \tau^*(|\rho'|^2 + |\tau'|^2) \exp[i(\varphi'_\rho + \varphi''_\rho)] = -\tau'$$

$$\rightarrow \quad |\tau| \exp[i(\varphi'_\rho + \varphi''_\rho)] = |\tau'| \exp[i(\varphi_\tau + \varphi'_\tau \pm \pi)]. \tag{8}$$

We conclude that $|\tau| = |\tau'| = |\tau''|$ and $|\rho| = |\rho'| = |\rho''|$. Moreover, a comparison with Eq.(4) reveals that $\varphi_\tau = \varphi''_\tau$ and, consequently, $\tau'' = \tau$. Substitution into Eq.(6) now yields $\rho^*\rho' = 1 - |\tau|^2$ and, therefore, $\rho' = \rho$. We also have, from Eq.(7), $\varphi'_\tau - \varphi_\rho = \varphi''_\rho - \varphi_\tau \pm \pi$, or, equivalently, $\varphi_\rho + \varphi''_\rho = \varphi_\tau + \varphi'_\tau \pm \pi$, but there is no new information in this equation. It is thus impossible with these techniques to relate $\varphi_\tau$ to $\varphi'_\tau$. (Bringing in the incident photon through channel 2 does not provide any new information either, as the preceding analysis has already revealed that the splitter's reflection and transmission coefficients for incidence from the backside are $\rho''' = \rho''$ and $\tau''' = \tau'$.)

If the splitter happens to have lateral structural symmetry (e.g., a multilayer stack, as opposed to a blazed grating), then we can say that $\varphi_\tau = \varphi'_\tau = \varphi''_\tau = \frac{1}{2}(\varphi_\rho + \varphi''_\rho) \pm \frac{1}{2}\pi$. Also, if there is symmetry between the front and back sides of the splitter, one can say that $\varphi_\rho = \varphi'_\rho = \varphi''_\rho$. In the absence of such symmetries, one must assume that, in general, $\varphi_\rho \neq \varphi''_\rho$ and $\varphi_\tau \neq \varphi'_\tau$.

**3. Number state $|n\rangle$ passing through a beam-splitter**. An insightful argument in *The Feynman Lectures on Physics* concerns the scattering of two or more indistinguishable Bose particles (e.g., identical photons) into a given quantum state.[11] Feynman explains why the probability of $n$ identical bosons piling up within a single number-state $|n\rangle$ is greater than the corresponding probability for distinguishable particles by a factor of $n!$. In the present section, we begin by extending Feynman's argument to the problem of $n$ identical photons in a number state $|n\rangle$ that, upon arriving at a beam-splitter, split into two groups of $m$ and $n - m$ photons. We then proceed to generalize the argument to the case of two (nearly) single-mode wavepackets $(\omega, \boldsymbol{k}_1, \hat{\boldsymbol{e}})$ and $(\omega, \boldsymbol{k}_2, \hat{\boldsymbol{e}})$, occupied by respective number-states $|n_1\rangle$ and $|n_2\rangle$, that enter a beam-splitter and subsequently produce wavepackets in the number-states $|m\rangle_3$ and $|n_1 + n_2 - m\rangle_4$ at the splitter's exit ports. (It is convenient to assume that the splitter has perpendicular arms and symmetric Fresnel coefficients for its various input/output ports, even though the analysis is readily extendible to the general case of a non-perpendicular and asymmetric splitter such as the one examined in the preceding section.)

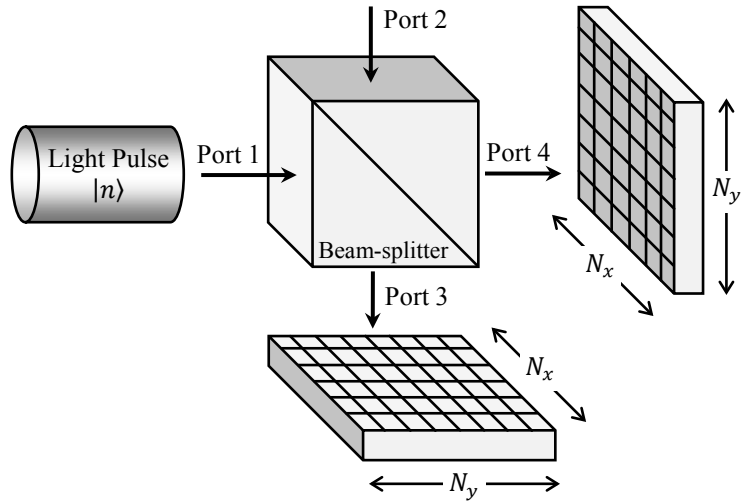

**Fig. 2**. A nearly single-mode light pulse arrives in the number state $|n\rangle$ at port 1 of a conventional beam-splitter whose Fresnel reflection and transmission coefficients are $\rho$ and $\tau$. Each photodetector array is finely divided in the beam's cross-sectional plane, and also in the time domain, so that a single-photon detection event can be attributed to a small area $\Delta x \Delta y$ in the cross-sectional $xy$ plane, and also to a short time interval $\Delta t$.

As shown in Fig.2, a nearly single-mode light pulse (i.e., a large wavepacket) in the number-state $|n\rangle$ enters port 1 of a conventional beam-splitter. The wavepacket is assumed to have a large cross-sectional area (compared to $\lambda^2$, where $\lambda$ is the free-space wavelength) and a long duration (compared to $2\pi/\omega$), so that it is a good approximation to a traveling monochromatic plane-wave; this is needed for the incoming light pulse to be considered "single-mode." The splitter and the detector arrays are taken to be sufficiently large to fully accommodate such an incoming beam. The amplitude reflection and transmission coefficients of the splitter are $\rho$ and $\tau$, with $|\rho|^2 + |\tau|^2 = 1$ and $\phi_\rho - \phi_\tau = \pm 90°$. The splitter is assumed to be symmetric, with $\rho$ and $\tau$ being the Fresnel coefficients for both ports 1 and 2.[13] The photodetector arrays in ports 3 and 4 are finely divided in the beam's cross-sectional plane, and also in the time domain, so that a single-photon detection event can be attributed to a small area $\Delta x \Delta y$ in the cross-sectional plane, and also to a short time interval $\Delta t$. The spatiotemporal dimensions $\Delta x$, $\Delta y$, $\Delta t$ of all detection cells must be as small as possible to enable unambiguous localization of individual photodetection events to the extent allowed by physical laws.[14] The total number of cells associated with each detector array in space-time is thus given by $N = N_x N_y N_t$; given the large volume of the single-mode light pulse(s) under consideration, the number $N$ of detection cells will be assumed to be far greater than the total number of the incoming photons. Detection of all $n$ photons within a specific set of $n$ cells from the two detector arrays constitutes a complete detection event.

In the following steps, we invoke Feynman's reasoning to calculate the number of ways that photons can be reflected and transmitted at the beam-splitter.

i) Assume at first that the $n$ photons of incoming light are distinguishable — as would be the case, for instance, if they were small balls of differing colors. The probability amplitude for $m$ photons to reflect while $n - m$ photons transmit at the splitter is $\rho^m \tau^{n-m}$.

ii) Each detector cell in space-time has a probability $1/N$ of picking up an incoming photon, with the corresponding probability amplitude being $1/\sqrt{N}$. The amplitude associated with $m$ photons arriving at $m$ specific cells in port 3, and the remaining photons arriving at $n - m$ specific cells in port 4, is thus given by $1/(\sqrt{N^m}\sqrt{N^{n-m}}) = 1/\sqrt{N^n}$. Multiplying this and the amplitude given in step (i), we obtain $\rho^m \tau^{n-m}/\sqrt{N^n}$ for the amplitude of each complete detection event.

iii) Now we take into account the indistinguishability of photons. For each specific set of detection cells, there exist $n!$ identical probability amplitudes which must be added together. This yields $n!\,\rho^m \tau^{n-m}/\sqrt{N^n}$ for the total amplitude of such an event.

iv) There exist $\binom{N}{m}$ distinct combinations of detection cells in port 3 and, similarly, $\binom{N}{n-m}$ distinct combinations of cells in port 4. Given that $N \gg n \geq m$, the following approximation is well justified:

$$\binom{N}{m} = \frac{N!}{m!(N-m)!} = \frac{N(N-1)(N-2)\cdots(N-m+1)}{m!}$$

$$= \frac{N^m}{m!}\left(1 - \frac{1}{N}\right)\left(1 - \frac{2}{N}\right)\cdots\left(1 - \frac{m-1}{N}\right)$$

$$\cong \frac{N^m}{m!} e^{-1/N} e^{-2/N} \cdots e^{-(m-1)/N}$$

$$= \frac{N^m}{m!} e^{-m(m-1)/(2N)} \cong \frac{N^m}{m!}. \tag{9}$$

A similar approximation, of course, can be used for $\binom{N}{n-m}$. The total number of ways in which $m$ photons get picked up at $m$ cells in port 3 while the remaining $n - m$ are picked up at $n - m$ cells in port 4 is, to an excellent approximation, given by $N^n/[m!\,(n - m)!]$. Considering that these detection events are naturally distinguishable, it is their probabilities (rather than their amplitudes) that must be added together. This means that, to determine the overall amplitude of detecting $m$ photons in port 3 (and $n - m$ in port 4), one must now multiply the amplitude $n!\,\rho^m \tau^{n-m}/\sqrt{N^n}$ derived in step (iii) with the *square root* of the number of distinct detection events, as follows:

$$\frac{n!\,\rho^m \tau^{n-m}}{\sqrt{m!(n-m)!}} = \sqrt{n!}\binom{n}{m}^{1/2} \rho^m \tau^{n-m}. \tag{10}$$

v) The above expression of the probability amplitude for detecting $m$ photons in port 3 (and the remaining $n - m$ in port 4) is almost complete except that it needs to be normalized by $\sqrt{n!}$. To appreciate the rationale for normalization, note that the corresponding probability of detecting $m$ photons in port 3, namely, $n!\binom{n}{m}|\rho|^{2m}|\tau|^{2(n-m)}$, when summed over $m$ (from 0 to $n$), yields $n!\,(|\rho|^2 + |\tau|^2)^n = n!$ instead of the required value of 1.0, hence the necessity of normalization.

A different justification for the need to normalize the probability amplitude of Eq.(10) by $\sqrt{n!}$ is based on the detection amplitude for the incoming light in the absence of the beam-splitter. Upon removing the beam-splitter, the light pulse will be fully detected by the detector array in port 4. Each detection event will then have amplitude $1/\sqrt{N^n}$, which must be multiplied by $n!$ to account for the indistinguishability of the photons. The total number of distinct configurations of $n$ detection cells in port 4 is now $\binom{N}{n} \cong N^n/n!$, whose square root multiplied into $n!/\sqrt{N^n}$ yields the overall amplitude for detecting all the incoming photons at port 4 as $\sqrt{n!}$. Considering that the probability of detecting all $n$ photons in this simple thought experiment should be 1.0, it is obvious that the results of such calculations must be normalized by $\sqrt{n!}$. We conclude that the amplitude for detecting $m$ photons in port 3 (and the remaining $n - m$ in port 4) is the normalized version of Eq.(10), namely, $\binom{n}{m}^{1/2} \rho^m \tau^{n-m}$.

Our elaborate five-step procedure for computing the probability amplitudes of various photon numbers emerging at the exit ports of a beam-splitter has now yielded a seemingly trivial result that is the same as what one obtains if the $n$ photons entering through port 1 are all distinguishable from one another. The power of the method, however, is revealed when the same procedure is applied to the case of two wavepackets in respective number-states $|n_1\rangle$ and $|n_2\rangle$ entering through the input ports 1 and 2 of the beam-splitter. This is due to the fact that quantum interference turns out to be far more consequential in the latter case. The case of two wavepackets simultaneously arriving at input ports 1 and 2 will be discussed after the following examples that pertain to the simpler case of a single wavepacket in the number-state $|n\rangle$ entering through port 1, as depicted in Fig.2.

**Example 1**. Let $\rho = 1/\sqrt{2}$ and $\tau = i/\sqrt{2}$. For $n = 2$, the probabilities of $m$ being 0, 1, or 2 will be ¼, ½, ¼, respectively. Similarly, for $n = 3$, the probabilities of $m = 0 - 3$ will be ⅛, ⅜, ⅜, ⅛.

**Example 2**. Let $n \gg 1$ and $|\tau| \gg |\rho|$. The probability of detecting $m$ photons in port 3 will be

$$\binom{n}{m}|\rho|^{2m}|\tau|^{2(n-m)} = |\tau|^{2n} \frac{n!}{m!(n-m)!}\left|\frac{\rho}{\tau}\right|^{2m}$$

$$= (1 - |\rho|^2)^n (n^m/m!)\left(1 - \tfrac{1}{n}\right)\left(1 - \tfrac{2}{n}\right)\cdots\left(1 - \tfrac{m-1}{n}\right)|\rho/\tau|^{2m}$$

$$\cong e^{-(\sqrt{n}|\rho|)^2} e^{-m(m-1)/(2n)} \frac{(\sqrt{n}|\rho/\tau|)^{2m}}{m!}$$

$$\cong e^{-(\sqrt{n}|\rho/\tau|)^2} \frac{(\sqrt{n}|\rho/\tau|)^{2m}}{m!}; \qquad m \lesssim \sqrt{n}. \tag{11}$$

Recalling that the single-mode quasi-classical (i.e., Glauber) coherent state is defined as[2,3] $|\gamma\rangle = e^{-\frac{1}{2}|\gamma|^2} \sum_{m=0}^{\infty} (\gamma^m/\sqrt{m!})|m\rangle$, it is seen that the packet exiting port 3 approximates such a beam with $\gamma = \sqrt{n}\rho/\tau$.

The preceding analysis parallels Feynman's "sum over alternative paths that could lead to an event," as described in his QED book.[15] It is readily extended to the more general case of two light pulses, one in the number state $|n_1\rangle$, the other in $|n_2\rangle$, that simultaneously arrive at ports 1 and 2 of a lossless beam-splitter. Assuming at first that all the photons are distinguishable, we take $m_1$ photons from port 1 and $m_2$ from port 2 to go into port 3, with the remaining photons then emerging at port 4. For a specific choice of these $m_1$ and $m_2$ photons, the corresponding probability amplitude at the beam-splitter is $\rho^{m_1} \tau^{m_2} \rho^{n_2-m_2} \tau^{n_1-m_1}$. Given that the number $N$ of detection cells at each port is very large (i.e., $N \gg n_1 + n_2$), the probability of detection at a specific set of $m_1 + m_2$ cells in port 3 is $1/N^{m_1+m_2}$; similarly, the probability of detection at a specific set of $(n_1 + n_2) - (m_1 + m_2)$ cells at port 4 is $1/N^{n_1+n_2-m_1-m_2}$. The product of these two probabilities is $1/N^{n_1+n_2}$, whose square root must be taken as the probability amplitude for the photons being picked up at those specific cells in ports 3 and 4. Thus, the overall amplitude for the selected $m_1 + m_2$ photons being picked up at a specific set of cells in port 3 while the remaining $(n_1 + n_2) - (m_1 + m_2)$ photons are picked up at a specific set of cells in port 4 is

$$\frac{\rho^{n_2+m_1-m_2} \tau^{n_1-m_1+m_2}}{\sqrt{N^{n_1+n_2}}}. \tag{12}$$

Now, there are $\binom{n_1}{m_1}$ ways to choose $m_1$ photons out of the $n_1$ that enter port 1, and $\binom{n_2}{m_2}$ ways to choose $m_2$ photons out of the $n_2$ that enter port 2. These photons can be distributed in $(m_1 + m_2)!$ distinct ways among the $m_1 + m_2$ specifically chosen cells at port 3. In similar fashion, there exist $(n_1 + n_2 - m_1 - m_2)!$ distinct ways of distributing the remaining photons among the chosen cells in port 4. All in all, the number of alternative arrangements for detecting $m_1 + m_2$ photons in specific cells in port 3, and $(n_1 + n_2) - (m_1 + m_2)$ photons in specific cells in port 4, turns out to be

$$\binom{n_1}{m_1} \binom{n_2}{m_2} (m_1 + m_2)! \, (n_1 + n_2 - m_1 - m_2)!. \tag{13}$$

We now invoke the fact that our $n_1 + n_2$ photons are indistinguishable, and that the probability amplitudes of Eq.(12), which are the same for all the various arrangements enumerated in Eq.(13), must be added together to arrive at the amplitude that our $m_1 + m_2$ photons are detected in a specific set of cells in port 3 while the remaining $(n_1 + n_2) - (m_1 + m_2)$ photons are detected in a specific set of cells in port 4. Moreover, when these specific detection events occur, there is no way to know the values of $m_1$ and $m_2$ separately; only their sum $m = m_1 + m_2$ will be known. Thus, for the specific set of cells under consideration, the total probability amplitude will be

$$\sum_{\substack{m_1=0 \\ m_1+m_2=m}}^{n_1} \sum_{m_2=0}^{n_2} \binom{n_1}{m_1} \binom{n_2}{m_2} (m_1 + m_2)! \, (n_1 + n_2 - m_1 - m_2)! \frac{\rho^{n_2+m_1-m_2} \tau^{n_1-m_1+m_2}}{\sqrt{N^{n_1+n_2}}}. \tag{14}$$

There exist $\binom{N}{m_1+m_2}$ different ways of picking $m = m_1 + m_2$ detection cells in port 3 and $\binom{N}{n_1+n_2-m_1-m_2}$ different ways of picking $n_1 + n_2 - m$ cells in port 4. Given that $N \gg n_1 + n_2$, the total number of ways in which $m = m_1 + m_2$ photons get picked up at the cells in port 3 while the remaining $n_1 + n_2 - m$ photons get picked up at those in port 4 is well approximated as

$$\binom{N}{m_1 + m_2}\binom{N}{n_1 + n_2 - m_1 - m_2} \cong \frac{N^{n_1+n_2}}{(m_1+m_2)!\,(n_1+n_2-m_1-m_2)!}. \tag{15}$$

Since these different configurations of our detector cells are naturally distinguishable, it is the probabilities (rather than their amplitudes) that must be added together, which means that the probability amplitude in Eq.(14) must be multiplied by the *square root* of the number of distinct configurations of detection cells given by Eq.(15). The resulting probability amplitude for detecting $m = m_1 + m_2$ photons in port 3 and the remaining $n_1 + n_2 - m$ photons in port 4 now becomes

$$\sum_{\substack{m_1=0 \\ m_1+m_2=m}}^{n_1} \sum_{m_2=0}^{n_2} \frac{n_1!\,n_2!\,\sqrt{(m_1+m_2)!(n_1+n_2-m_1-m_2)!}\,\rho^{n_2+m_1-m_2}\tau^{n_1-m_1+m_2}}{m_1!\,m_2!\,(n_1-m_1)!\,(n_2-m_2)!}. \tag{16}$$

We still need to normalize the above amplitude by $\sqrt{n_1!\,n_2!}$, since, upon removing the beam-splitter, the amplitude for directly detecting all $n_1$ photons of the first light pulse in port 4 is found to be $\sqrt{n_1!}$ and, similarly, the amplitude for directly detecting all $n_2$ photons of the second light pulse in port 3 turns out to be $\sqrt{n_2!}$. The properly normalized amplitude for detecting $m = m_1 + m_2$ photons in port 3 and the remaining $n_1 + n_2 - m$ photons in port 4 may now be written as follows:

$$\sum_{\substack{m_1=0 \\ m_1+m_2=m}}^{n_1} \sum_{m_2=0}^{n_2} \frac{\sqrt{n_1!\,n_2!\,(m_1+m_2)!\,(n_1+n_2-m_1-m_2)!}\,\rho^{n_2+m_1-m_2}\tau^{n_1-m_1+m_2}}{m_1!\,m_2!\,(n_1-m_1)!\,(n_2-m_2)!}. \tag{17}$$

The expression in Eq.(17) is precisely the same as that obtained by applying the operator algebra[1,2] to the system depicted in Fig.2, when light pulses in the number states $|n_1\rangle$ and $|n_2\rangle$ simultaneously arrive at ports 1 and 2. A brief analysis of this problem with the aid of operator algebra is given in the next section.

**Digression**: The following streamlined version of Eq.(17) may be easier to memorize, and might even provide an incentive to search for an alternative explanation (or plausibility argument) for the probability amplitude of detecting $m$ photons in port 3 and $n_1 + n_2 - m$ photons in port 4:

$$\sum_{\substack{m_1=0 \\ m_1+m_2=m}}^{n_1} \sum_{m_2=0}^{n_2} \sqrt{\binom{n_1}{m_1}\binom{n_2}{m_2}\binom{m_1+m_2}{m_1}\binom{n_1+n_2-m_1-m_2}{n_1-m_1}}\,\rho^{n_2+m_1-m_2}\tau^{n_1-m_1+m_2}. \tag{18}$$

Unfortunately, try as we might, we have failed to find an alternative physical justification for the relatively simple structure of the square root factor that appears in the above equation.

**4. Analysis of a beam-splitter using operator algebra**. In quantum optics, the standard approach to analyzing the action of lossless beam-splitters on incoming photon-number states relies on the fundamental properties of the annihilation and creation operators $\hat{a}$ and $\hat{a}^\dagger$.[1-4] Consider a beam-splitter whose amplitude reflection and transmission coefficients satisfy $|\rho|^2 + |\tau|^2 = 1$ and $\phi_\rho - \phi_\tau = \pm 90°$. The splitter is assumed to be symmetric, with $\rho$ and $\tau$ representing the Fresnel coefficients for both input ports 1 and 2. Starting with the $|0\rangle_1|0\rangle_2$ vacuum mode in the splitter's input space (ports 1 and 2), creation operators generate the number state $|n_1\rangle_1|n_2\rangle_2$ in the input space; that is,

$$|n_1\rangle_1|n_2\rangle_2 = \frac{(\hat{a}_1^\dagger)^{n_1}(\hat{a}_2^\dagger)^{n_2}}{\sqrt{n_1!\,n_2!}}|0\rangle_1|0\rangle_2. \tag{19}$$

In the output space (ports 3 and 4), the annihilation and creation operators are given by $\hat{a}_3 = \rho\hat{a}_1 + \tau\hat{a}_2$, $\hat{a}_4 = \tau\hat{a}_1 + \rho\hat{a}_2$, $\hat{a}_3^\dagger = \rho^*\hat{a}_1^\dagger + \tau^*\hat{a}_2^\dagger$, and $\hat{a}_4^\dagger = \tau^*\hat{a}_1^\dagger + \rho^*\hat{a}_2^\dagger$. It is straightforward to show that $[\hat{a}_3, \hat{a}_3^\dagger] = 1$, $[\hat{a}_4, \hat{a}_4^\dagger] = 1$, and $[\hat{a}_3, \hat{a}_4] = [\hat{a}_3^\dagger, \hat{a}_4^\dagger] = [\hat{a}_3, \hat{a}_4^\dagger] = [\hat{a}_3^\dagger, \hat{a}_4] = 0$. Solving the above equations for the input operators $\hat{a}_1^\dagger$, $\hat{a}_2^\dagger$ in terms of the output operators $\hat{a}_3^\dagger$, $\hat{a}_4^\dagger$, one finds

$$\hat{a}_1^\dagger = \rho\hat{a}_3^\dagger + \tau\hat{a}_4^\dagger, \tag{20}$$

$$\hat{a}_2^\dagger = \tau\hat{a}_3^\dagger + \rho\hat{a}_4^\dagger. \tag{21}$$

Substitution into the right-hand side of Eq.(19) then yields

$$\frac{(\hat{a}_1^\dagger)^{n_1}(\hat{a}_2^\dagger)^{n_2}}{\sqrt{n_1!n_2!}}|0\rangle_1|0\rangle_2 = \frac{(\rho\hat{a}_3^\dagger + \tau\hat{a}_4^\dagger)^{n_1}(\tau\hat{a}_3^\dagger + \rho\hat{a}_4^\dagger)^{n_2}}{\sqrt{n_1!n_2!}}|0\rangle_1|0\rangle_2$$

$$= \frac{\sum_{m_1=0}^{n_1}\binom{n_1}{m_1}\rho^{m_1}\tau^{(n_1-m_1)}(\hat{a}_3^\dagger)^{m_1}(\hat{a}_4^\dagger)^{(n_1-m_1)}\sum_{m_2=0}^{n_2}\binom{n_2}{m_2}\tau^{m_2}\rho^{(n_2-m_2)}(\hat{a}_3^\dagger)^{m_2}(\hat{a}_4^\dagger)^{(n_2-m_2)}}{\sqrt{n_1!n_2!}}|0\rangle_1|0\rangle_2$$

$$= \frac{\sum_{m_1=0}^{n_1}\sum_{m_2=0}^{n_2}\binom{n_1}{m_1}\binom{n_2}{m_2}\rho^{(n_2+m_1-m_2)}\tau^{(n_1-m_1+m_2)}(\hat{a}_3^\dagger)^{m_1+m_2}(\hat{a}_4^\dagger)^{n_1+n_2-m_1-m_2}}{\sqrt{n_1!n_2!}}|0\rangle_1|0\rangle_2$$

$$= \sum_{m_1=0}^{n_1}\sum_{m_2=0}^{n_2}\frac{\sqrt{n_1!n_2!(m_1+m_2)!(n_1+n_2-m_1-m_2)!}}{m_1!m_2!(n_1-m_1)!(n_2-m_2)!}\rho^{(n_2+m_1-m_2)}\tau^{(n_1-m_1+m_2)}|m_1+m_2\rangle_3|n_1+n_2-m_1-m_2\rangle_4. \tag{22}$$

The last expression contains the probability amplitudes for the various number states $|m_3\rangle_3|m_4 = n_1 + n_2 - m_3\rangle_4$ emerging at ports 3 and 4. Note that, for each value of $m_3$, one must add up all the probability amplitudes associated with terms $m_1$ and $m_2$ whose sum equals $m_3$.

---

**Example 3**. Let $n_2 = 0$. The probability amplitude of $|m\rangle_3|n_1 - m\rangle_4$ will be $\binom{n_1}{m}^{1/2}\rho^m\tau^{n_1-m}$. Stated differently, the probability that $m$ photons (out of $n_1$ incident) are reflected is $\binom{n_1}{m}|\rho|^{2m}|\tau|^{2(n_1-m)}$.

---

**Example 4**. Let $n_1 = n_2 = 1$. The probability amplitudes of the output states $|0\rangle_3|2\rangle_4$, $|1\rangle_3|1\rangle_4$, and $|2\rangle_3|0\rangle_4$ will then be $\sqrt{2}\rho\tau$, $\rho^2 + \tau^2$, and $\sqrt{2}\rho\tau$, respectively. Considering that $|\rho^2 + \tau^2| = |\rho|^2 - |\tau|^2$, it is clear that the three probabilities will add up to 1. Also, when $|\rho| = |\tau| = 1/\sqrt{2}$, the probability of the output being in the $|1\rangle_3|1\rangle_4$ state vanishes. The two input photons will then coalesce, appearing together either in port 3 or in port 4 with equal probabilities. This, of course, is the well-known Hong-Ou-Mandel effect.[16]

Note that, in the special case of $\rho = 1/\sqrt{2}$ and $\tau = i/\sqrt{2}$, the appearance of a single photon in each exit port requires that both incoming photons be reflected, or both be transmitted, at the splitter. These two events have respective probability amplitudes $\rho^2 = \frac{1}{2}$ and $\tau^2 = (i/\sqrt{2})^2 = -\frac{1}{2}$. Since the two events are indistinguishable, their probability amplitudes must be added together, which results in zero probability for the appearance of a single photon in each exit port. As for the 2-photon coalescence event in port 3 (or port 4), the splitter must reflect one photon and transmit the other. The amplitude for reflection at one input port and transmission at the other is $\rho\tau = i/2$. Considering that $|\rho\tau|^2 = \frac{1}{4}$, it is seen that the probabilities of the three possible outcomes in this example fail to add up to 1 (that is, $\frac{1}{4} + \frac{1}{4} + 0 \neq 1$) unless the relevant Bose enhancement factor (i.e., the radical in Eq.(18)) for each of the 2-photon coalescence events is properly accounted for.

It is in simple examples such as this that we find a clear need for invoking the arguments of Sec.3, which account for all allowed distinguishable as well as indistinguishable detection events. Even though, in practice, the detectors may not be subdivided into a multitude of detection cells, the theoretical possibility of such subdivisions (in both time and space) must be carefully considered since, as a matter of principle, the distinguishable and indistinguishable events require different treatments.

---

**Example 5**. Let $n_1 = n$, $n_2 = 0$, and $m_3 = m_1 + m_2 = 1$. The probability amplitude for detecting exactly one photon in port 3 and $n - 1$ photons in port 4 is $\sqrt{n}\rho\tau^{n-1}$. Now, for a given $n$, it is possible to choose $\rho$ so small as to make the magnitude of $\tau^{n-1}$ arbitrarily close to unity. The above probability amplitude will then be $\cong \sqrt{n}\rho$ (aside from a possible phase factor). It is thus clear that the probability amplitude for single-photon *annihilation* is proportional to $\sqrt{n}$. (The proportionality constant $\rho$ is obviously needed here, since it is the amplitude for a single photon to leave the incoming $n$-photon packet via reflection at the beam-splitter.)

In the same setting, with $\rho$ being sufficiently small, the probability amplitude for 2-photon reflection at the splitter is $\cong \sqrt{n(n-1)/2}\,\rho^2$, which is off by a factor of $\sqrt{2}$ when compared against the operator identity $\hat{a}^2|n\rangle = \sqrt{n(n-1)}|n-2\rangle$. The discrepancy is due to the fact that the splitter in this instance no longer acts as an elementary 2-photon annihilator, simply because the emergence of $|n-2\rangle$ in port 3 is accompanied by the simultaneous appearance of the 2-photon state $|2\rangle$ in port 4. A properly constructed 2-photon annihilator would consist of a pair of back-to-back splitters (each having a sufficiently small reflection coefficient $\rho$), so that transmission of an $n$-photon wavepacket through the first splitter results in the loss of a single photon, followed by the loss of a second photon upon passage through the second splitter. With the caveat that a single photon has appeared in port 3 of each splitter, the state of the wavepacket emerging at the second splitter's port 4 will now be $\cong \sqrt{n(n-1)}\rho^2|n-2\rangle$.

The artificial device of a weak beam-splitter followed by a reliable photodetector used in the present example is convenient, of course, but by no means necessary. A single atom in its ground state, capable of absorbing a photon of energy $\hbar\omega$ from the single-mode number state $|n\rangle$, will also exhibit an absorption probability amplitude proportional to $\sqrt{n}$. This is because each incident photon has a $1/n$ probability (corresponding to a $1/\sqrt{n}$ amplitude) of being absorbed by the atom. Since the photons are indistinguishable, their absorption amplitudes must be added together, yielding an overall probability amplitude of $n/\sqrt{n} = \sqrt{n}$ for single-photon annihilation.

---

**Example 6**. Let $n_1 = n$, $n_2 = 1$, and $m_3 = m_1 + m_2 = 0$. The probability amplitude for detecting exactly $n + 1$ photons in port 4 is $\sqrt{n+1}\rho\tau^n$. Once again, for a given $n$, it is possible to choose $\rho$ so small as to make the magnitude of $\tau^n$ arbitrarily close to unity. The probability amplitude, aside from a possible phase factor, will then be $\cong \sqrt{n+1}\rho$. The amplitude for single-photon *creation* is thus seen to be proportional to $\sqrt{n+1}$. (The proportionality constant $\rho$ is needed here, since it is the amplitude for the single photon arriving at port 2 to bounce off the splitter and merge with the $n$-photon packet that enters through port 1 and goes directly to port 4.)

In analogy with the preceding example, consider a single atom in an excited state, capable of emitting a photon of energy $\hbar\omega$ into the number state $|n\rangle$ of a single-mode wavepacket that engulfs the atom. The probability amplitude that this emission will raise the number-state to $|n + 1\rangle$ is proportional to $\sqrt{n+1}$, ostensibly due to the possibility of stimulation of the atom by any one of the $n$ incoming photons or by the corresponding vacuum mode.[17] These $n + 1$ possibilities, each with its probability amplitude of $1/\sqrt{n+1}$, are indistinguishable and, therefore, their amplitudes must be added together, yielding an overall probability amplitude of $\sqrt{n+1}$ for single-photon creation.

---

**5. Concluding remarks**. The classical Fresnel reflection and transmission coefficients $\rho$ and $\tau$ that relate the complex $E$-field amplitudes of the incident, reflected, and transmitted beams at the input and output ports of a lossless beam-splitter can also serve as probability amplitudes for single-photon reflection and transmission at the corresponding ports. We have shown that, in general, while $|\rho|$ (and also $|\tau|$) must be the same for different available input/output ports—satisfying $|\rho|^2 + |\tau|^2 = 1$ in every instance—the phase angles $\varphi_\rho$ associated with different pairs of input/output ports could differ from one another, as can the phase angles $\varphi_\tau$. Nevertheless, the mandatory relations existing among the various phase angles ensure the universality of the essential properties of beam-splitters. For most practical applications where the splitters are designed to have lateral symmetry as well as symmetry between their front and back sides, the relation $\varphi_\rho - \varphi_\tau = \pm 90°$ will hold, with all pairs of input/output ports sharing the same $\varphi_\rho$ and also the same $\varphi_\tau$.

The main goal of the paper has been to derive the probability amplitudes for different photon-number states that emerge at the output ports of a lossless beam-splitter when wavepackets in the number states $|n_1\rangle$ and $|n_2\rangle$ simultaneously arrive at its input ports. We used an extension of the argument originally made by Feynman in his scattering analysis of indistinguishable Bose particles to arrive at our final result given in Eq.(17). We also showed how the same result is obtained using standard methods of quantum optics that rely on the properties of annihilation and creation operators $\hat{a}$ and $\hat{a}^\dagger$; see Eq.(22). The operator-based methods, of course, are powerful and flexible and can handle complex problems beyond the reach of our elementary arguments. For example, one can use the operator algebra to prove that two wavepackets in the Glauber coherent states $|\gamma_1\rangle$ and $|\gamma_2\rangle$ entering a lossless beam-splitter will emerge as coherent states $|\rho\gamma_1 + \tau\gamma_2\rangle$ and $|\tau\gamma_1 + \rho\gamma_2\rangle$ at the corresponding exit ports.[18] This and numerous other examples notwithstanding, we believe that, for pedagogical purposes, it is worthwhile seeing that the very important property of beam-splitters given by Eq.(22) can be "explained" at a more elementary level using the simple addition of probability amplitudes associated with indistinguishable events.

The addition of probability amplitudes was also invoked in Examples 5 and 6 to elucidate the Bose enhancement implicit in the standard formulas $\hat{a}|n\rangle = \sqrt{n}|n-1\rangle$ and $\hat{a}^\dagger|n\rangle = \sqrt{n+1}|n+1\rangle$. In the presence of a single-mode wavepacket in the number-state $|n\rangle$, the resonant absorption of a photon by an atom in its ground state, or the stimulated emission of a photon by an atom in an excited state, could serve as simple examples to corroborate the action of annihilation and creation operators via a straightforward application of the Feynman method.

The feasibility of detecting the individual photons of a wavepacket at specific locations in space and time served as a basis for our method of analyzing the behavior of a beam-splitter that culminated in Eq.(17). While detector arrays capable of localizing individual photons in space and time are commercially available nowadays, it is not known whether the corresponding technologies have reached their physical limits, or if more accurate detector arrays with smaller pixels, narrower dead-zones, less jitter, and shorter reset times will become available in the future. It is intriguing to contemplate the fundamental limitations that existing (or future) photodetector arrays impose on the minimum detector area $\Delta x \Delta y$ and also on the minimum detection time $\Delta t$. At a fundamental level, it might be useful to know, for instance, if pinpointing individual photons with smaller detector cells would require longer exposure times.

We close by speculating that the required normalization of Eq.(10) by $\sqrt{n!}$ (and of Eq.(16) by $\sqrt{n_1! n_2!}$) may be intimately tied to the action of the $n$-fold annihilation operator $\hat{a}^n$ on the number-state $|n\rangle$, which is formally expressed as $\hat{a}^n|n\rangle = \sqrt{n!}|0\rangle$. The process described in Sec.3, where all $n$ photons contained in a single-mode packet are absorbed by the detectors, is essentially an $n$-photon annihilation operation that should end up in the unnormalized vacuum state $\sqrt{n!}|0\rangle$. It is, therefore, plausible that the probability amplitudes obtained by adding up the amplitudes of all the indistinguishable events should require normalization by the corresponding Bose enhancement factor, namely, $\sqrt{n!}$. Similarly, in the case of two single-mode packets arriving in the number states $|n_1\rangle$ and $|n_2\rangle$ at ports 1 and 2, the final vacuum state (reached after detecting all the incoming photons) should be $\sqrt{n_1!n_2!}|0\rangle$, which necessitates that the various probability amplitudes evaluated at the exit ports be normalized by the corresponding enhancement factor $\sqrt{n_1!n_2!}$.

**Acknowledgement**. The authors are grateful to the editor and two anonymous referees whose detailed comments and valuable suggestions have helped to improve the overall clarity of our presentation.

**Conflict of Interest Statement**. The authors have no conflicts to disclose.